# Improved OMP Approach to Sparse Multi-path Channel Estimation via Adaptive Inter-atom Interference Mitigation

Ruiming Yang, Qun Wan, Yipeng Liu and Wanlin Yang
Department of Electronic Engineering
University of Electronic Science and Technology of China
Chengdu, China
{ shan99, wanqun, liuyipeng, wlyang}@uestc.edu.cn

*Abstract*—Since most components of sparse multi-path channel (SMPC) are zero, impulse response of SMPC can be recovered from a short training sequence. Though the ordinary orthogonal matching pursuit (OMP) algorithm provides a very fast implementation of SMPC estimation, it suffers from inter-atom interference (IAI), especially in the case of SMPC with a large delay spread and short training sequence. In this paper, an adaptive IAI mitigation method is proposed to improve the performance of SMPC estimation based on a general OMP algorithm. Unlike the ordinary OMP algorithm, a sensing dictionary is designed adaptively and posterior information is utilized efficiently to prevent false atoms from being selected due to serious IAI. Numeral experiments illustrate that the proposed general OMP algorithm based on adaptive IAI mitigation outperform both the ordinary OMP algorithm and the general OMP algorithm based on non-adaptive IAI mitigation.

## I. INTRODUCTION

In the last decade, how to overcome the scarcity of spectral resource to meet the ever-growing need for high data rate was a great challenge for communication engineers. One way to achieve a high data rate is to simply increase the transmission speed. Due to the time delay spread of multi-path channel, the channel impulse response easily spans several hundred symbol intervals. If a standard least-squares (LS) type channel estimator is used, current training sequence is generally short to provide accurate channel estimation [1]. Fortunately, sparse multi-path channel (SMPC) is frequently encountered in wireless communication applications, such as terrestrial transmission channel of high definition television (HDTV) signals [2], hilly terrain delay profile of multi-path in the broadband wireless communication [3] and typical underwater acoustic channels [4]. Among the large number of SMPC entries, only a small portion is significantly different from zero. Taking advantage of the sparsity, impulse response of SMPC can be recovered from relatively small number of received data and training data. However, finding the sparsest solution is an NP-Hard combinatorial problem.

In order to find a suboptimal but sufficient sparse solution, several greedy algorithms [5, 6] and optimization methods [7] have been proposed. Instead of just representing the received signal as accurate as possible by channel impulse response weighted superposition of the transmitted signals, they have available a redundant dictionary and their goal is to obtain not only accurate but also the sparsest possible representation of the received signal from that over-complete dictionary. Among these suboptimal methods, matching pursuit (MP) algorithm can provide a very fast implementation of sparse approximation [8]. It has been inspiring for many researchers and different variations of this algorithm were proposed. The most famous one is orthogonal matching pursuit (OMP) algorithm [9, 10]. Using OMP, the convergence problem in MP algorithm based on reselection of the atoms is eliminated. It was also verified that by avoiding the re-selection problem, more accurate channel estimates can be obtained by using the OMP algorithm [10]. However, according to the sufficient condition developed by Tropp [11], both the suboptimal algorithms (MP and OMP) suffer from inter-atom interference (IAI) due to coherency and redundancy of dictionary, especially in the case of SMPC with either large time delay spread or relatively small number of training data and received data.

Unlike the atoms corresponding to zero entries of SMPC which will not affect the estimated value of any entries of SMPC, the atoms corresponding to nonzero entries of SMPC will draw the estimated value of each entry of SMPC away from its correct value. As a result of serious IAI, we may either choose a false atom when the associated entry of SMPC is zero or omit a correct atom when the associated entry is nonzero. Recently, a general OMP algorithm was developed to improve the performance of the ordinary OMP algorithm in the case of highly coherent dictionary through introducing a sensing dictionary [12]. However, it only considered the noiseless situation and the sensing dictionary is non-adaptively designed, which is independent of the received data.

In this paper, a novel adaptive IAI mitigation method is proposed to improve the performance of SMPC estimation based on the general OMP algorithm. Unlike the non-adaptive

This work was supported in part by the National Natural Science Foundation of China under grant 60772146, the National High Technology Research and Development Program of China (863 Program) under grant 2008AA12Z306 and in part by Science Foundation of Ministry of Education of China under grant 109139.)

sensing dictionary used in the previous general OMP algorithm, an adaptively designed sensing dictionary is build up and posterior information is utilized efficiently to prevent false atoms from being selected due to serious IAI. Numeral experiments illustrate that the performance of the proposed general OMP algorithm based on adaptive IAI mitigation is better than that of both the ordinary OMP algorithm and the general OMP algorithm based on non-adaptive IAI mitigation.

This paper is organized as follows. In Section II, the sparse multi-path channel model is presented and inter-atom interference problem is formulated. The principle of inter-atom interference mitigation is given in Section III and the approach to inter-atom interference mitigation is given in Section IV. Finally, we compare the performance of the proposed algorithm with other algorithms via simulation over wireless Gauss channel in Section V and conclusions are given in Section VI.

*Notation:* In this paper, the superscript $T$ stands for transposition. Bold capital letters denote a matrix whereas bold small letters indicate a vector. $|\cdot|$ stands for the absolute value of a scalar or each component of a vector. Finally, $\hat{\mathbf{h}}$ and $\mathbf{h}$ indicate an estimate and a real value of vector $\mathbf{h}$, respectively.

## II. PROBLEM FORMULATION

Let's transmit the training sequence $s(n)$, $n = 0, 1, \cdots, N-1$, through a stationary multi-path sparse channel. The training sequence symbols $s(n)$ for $n < 0$ can be obtained from the previous estimates or for the first arriving frame they are assumed to be zero [13]. The received base-band signal samples can be modeled as

$$r_t = \sum_{i=0}^{L-1} s(t-i) h_i + e_t, \quad (1)$$

where $t = 0, 1, \cdots, N-1$, $h_i$ is the channel impulse response of length $L$, $e_t$ is additive white Gaussian noise with zero mean and variance $\sigma_e^2$. Denote the power of training sequence and the received signal by $\sigma_s^2$ and $\sigma_r^2$, respectively. In the vector form, we have

$$\mathbf{r} = \mathbf{S}\mathbf{h} + \mathbf{e}, \quad (2)$$

where $\mathbf{h} = \begin{bmatrix} h_0 & h_1 & \cdots & h_{L-1} \end{bmatrix}^T$, $\mathbf{r} = \begin{bmatrix} r_0 & r_1 & \cdots & r_{N-1} \end{bmatrix}^T$, $\mathbf{e} = \begin{bmatrix} e_0 & e_1 & \cdots & e_{N-1} \end{bmatrix}^T$ and $\mathbf{S}$ is the known training matrix given by

$$\mathbf{S} = \begin{bmatrix} s(0) & s(-1) & \ldots & s(-L+1) \\ s(1) & s(0) & \ldots & s(1) \\ \vdots & \vdots & \ddots & \vdots \\ s(N-1) & s(N-2) & \ldots & s(N-L) \end{bmatrix} = \begin{bmatrix} \mathbf{s}_0 & \mathbf{s}_1 & \cdots & \mathbf{s}_{L-1} \end{bmatrix}, \quad (3)$$

Denote the number of nonzero entries of $\mathbf{h}$ as $K$. The channel $\mathbf{h}$ is sparse if $K \ll L$ is satisfied. In the context of sparse analysis, $\mathbf{S}$ is called dictionary and the column vector $\mathbf{s}_i$ is called atom, $i = 0, 1, \cdots, L-1$. As a result of short length training sequence, which improves throughput efficiency for the systems where transmitted packet length is short, the dictionary is highly redundant. In other word, the dimension of the received base-band signal vector $\mathbf{r}$ is much smaller than the number of atoms in the dictionary, i.e., $N \ll L$.

Though the problem of finding the best sparse channel solution from the contaminated received signal is NP-Hard, suboptimal solutions may be sufficient in wireless communication [1]. Among these methods, OMP is an attractive algorithm since it is fast and easy to implement [8]. The ordinary OMP algorithm iteratively selects an atom in dictionary that correlates most strongly with the residual signal. At each step $k$, the best atom $\mathbf{s}_{m_k}$ is selected through the peak position searching as

$$m_k = \arg \max_{0 \leq i \leq L-1} \hat{h}_i^{(k)}, \quad (4)$$

$$\hat{\mathbf{h}}^{(k)} = \begin{bmatrix} \hat{h}_0^{(k)} & \hat{h}_1^{(k)} & \cdots & \hat{h}_{L-1}^{(k)} \end{bmatrix}^T = \left| \mathbf{S}^T \mathbf{g}_k \right|, \quad (5)$$

where $k = 0, 1, \cdots, K-1$. We have $\mathbf{g}_0 = \mathbf{r}$ for initialization and $\mathbf{g}_{k+1} = \mathbf{P}_k \mathbf{r}$ for k = 0, 1, ... , K-2; $\mathbf{P}_k = \mathbf{I}_M - \hat{\mathbf{A}}^{(k)} \left( \hat{\mathbf{A}}^{(k)T} \hat{\mathbf{A}}^{(k)} \right)^{-1} \hat{\mathbf{A}}^{(k)T}$, $\hat{\mathbf{A}}^{(k)} = \begin{bmatrix} \mathbf{s}_{m_0} & \mathbf{s}_{m_1} & \cdots & \mathbf{s}_{m_k} \end{bmatrix}$ and $\mathbf{I}_M$ is an identity matrix.

To illustrate the effect of IAI on the performance of OMP algorithm, e.g., at the initialization step, we express the sparse channel estimation as

$$\hat{\mathbf{h}}^{(0)} = \left| \mathbf{S}^T \mathbf{g}_0 \right| = \left| \mathbf{S}^T \mathbf{r} \right| = \left| \mathbf{S}^T (\mathbf{S}\mathbf{h} + \mathbf{e}) \right|, \quad (6)$$

or

$$\hat{h}_i^{(0)} = \left| \sum_{l=0}^{L-1} \mathbf{s}_i^T \mathbf{s}_l h_l + \mathbf{s}_i^T \mathbf{e} \right|, \quad (7)$$

for $i = 0, 1, \cdots, L-1$. If $h_l = 0$, the IAI item $\mathbf{s}_i^T \mathbf{s}_l$ can not affect the estimated value of $\hat{h}_i^{(0)}$. However, if $h_l \neq 0$, the IAI item $\mathbf{s}_i^T \mathbf{s}_l$ will draw the estimated value of $\hat{h}_i^{(0)}$ away from its correct value $h_i$. As a result, we may either choose a false atom when $h_i = 0$ or omit a correct atom when $h_i \neq 0$ at this step if IAI is large enough. Here, the problem is how to mitigate the effect of IAI on the performance of OMP algorithm.

## III. PRINCIPLE OF IAI MITIGATION

In order to identify the correct atoms in the case of high IAI level, we resort to the general OMP based on a sensing dictionary $\mathbf{W}$, and use $\hat{\mathbf{h}}^{(k)} = |\mathbf{W}^T \mathbf{g}_k|$ rather than $\hat{\mathbf{h}}^{(k)} = |\mathbf{S}^T \mathbf{g}_k|$ in (5). Obviously, the ordinary OMP is a special case of the general OMP with $\mathbf{W} = \mathbf{S}$. At the initialization step of the general OMP, e.g., we have

$$\hat{\mathbf{h}}^{(0)} = |\mathbf{W}^T \mathbf{g}_0| = |\mathbf{W}^T \mathbf{r}| = |\mathbf{W}^T (\mathbf{S}\mathbf{h} + \mathbf{e})|, \quad (8)$$

or

$$\hat{h}_i^{(0)} = \left| \sum_{l=0}^{L-1} \mathbf{w}_i^T \mathbf{s}_l h_l + \mathbf{w}_i^T \mathbf{e} \right|, \quad (9)$$

for $i = 0, 1, \cdots, L-1$.

If $h_l = 0$, the IAI item $\mathbf{w}_i^T \mathbf{s}_l$ can be ignored. However, if $h_l \neq 0$, the IAI item $\mathbf{w}_i^T \mathbf{s}_l$ should be as small as possible no matter $h_i = 0$ or $h_i \neq 0$. Otherwise, the IAI item $\mathbf{w}_i^T \mathbf{s}_l$ will draw the estimated value of $\hat{h}_i^{(0)}$ away from its correct value $h_i$. Thus, we may design each column vector of $\mathbf{W}$, i.e. the sensing vector $\mathbf{w}_i$, as the solution to the following minimum interference distortionless response (MIDR) problem:

$$\min_{\mathbf{w}_i} \mathbf{w}_i^T \mathbf{B} \mathbf{B}^T \mathbf{w}_i, \quad (10)$$

$$\text{s.t. } \mathbf{s}_i^H \mathbf{w}_i = 1, \quad (11)$$

where $\mathbf{B}$ consists of the correct atoms corresponding to the nonzero entries of sparse channel $\mathbf{h}$. The closed-form solution is given by

$$\mathbf{w}_i = \mathbf{Q}_i \mathbf{s}_i, \quad (12)$$

for $i = 0, 1, \cdots, L-1$, where

$$\mathbf{Q}_i = \frac{1}{\mathbf{s}_i^T (\mathbf{B}^T \mathbf{B} + \alpha \mathbf{I}_M)^{-1} \mathbf{s}_i} (\mathbf{B}^T \mathbf{B} + \alpha \mathbf{I}_M)^{-1}, \quad (13)$$

and $\alpha$ is a positive regularization parameter.

When $\mathbf{s}_i$ is one of the column vectors of $\mathbf{B}$, i.e., the correct atoms, the minimum variance condition (10) will mitigate the correlation between the corresponding sensing vector $\mathbf{w}_i$ and other correct atoms, while the distortionless response constraint (11) will maintain the correlation between $\mathbf{w}_i$ and this correct atom. As a result, the nonzero entries of $\mathbf{h}$ corresponding to the correct atoms are estimated with distortion as small as possible. On the other hand, when $\mathbf{s}_i$ is not one of the column vectors of $\mathbf{B}$, the minimum variance condition (10) will prevent false atoms being selected through mitigating the correlation between the corresponding sensing vector $\mathbf{w}_i$ and all the correct atoms.

However, the closed-form solution (12) is unrealistic because either $\mathbf{B}$ is not available or the correct atoms themselves are to be identified.

## IV. ADAPTIVE APPROACH TO IAI MITIGATION

Given the received signal $\mathbf{r}$, the probability of appearance in the reconstruction of $\mathbf{r}$ is different for different atom [14]. Like the ordinary OMP algorithm, we take the correlation between the received vector (or the residual vector) and each atom in the dictionary as an approximate measure of this probability. Unlike the ordinary OMP algorithm, which only uses this measure to select best atom sequentially, we exploit it to design an adaptive sensing dictionary based on the following approximation:

$$\mathbf{B}\mathbf{B}^T \approx \mathbf{S}\mathbf{U}^{(k)}\mathbf{S}^T, \quad (14)$$

where

$$\mathbf{U}^{(k)} = \text{diag}(|\hat{\mathbf{h}}^{(k)}|^\rho), \quad (15)$$

$$\hat{\mathbf{h}}^{(k)} = |\mathbf{W}^T \mathbf{g}_k|, \quad (16)$$

and $\rho > 0$. Substituting (14) into (10) yields the adaptive sensing vector as the solution to the following approximate MIDR optimization problem:

$$\min_{\mathbf{w}_i} \mathbf{w}_i^T \mathbf{B} \mathbf{B}^T \mathbf{w}_i, \quad (17)$$

$$\text{s.t. } \mathbf{s}_i^H \mathbf{w}_i = 1. \quad (18)$$

Similarly, the closed-form solution can be given by

$$\mathbf{w}_i = \mathbf{D}_i \mathbf{s}_i, \quad (19)$$

where

$$\mathbf{D}_i = \frac{1}{\mathbf{s}_i^T (\mathbf{S}\mathbf{U}^{(k)}\mathbf{S}^T + \beta \mathbf{I}_M)^{-1} \mathbf{s}_i} (\mathbf{S}\mathbf{U}^{(k)}\mathbf{S}^T + \beta \mathbf{I}_M)^{-1}, \quad (20)$$

for $i = 0, 1, \cdots, L-1$, and $\beta$ is a positive regularization parameter. Because $\mathbf{U}^{(k)}$ in (20) is calculated from the sensing dictionary itself, we must set an initial sensing dictionary such as $\mathbf{W} = \mathbf{S}$.

The advantage of the sensing dictionary given by (19) is the adaptive function of IAI mitigation as a result of both the adaptive minimum interference optimization and the distortionless response constraint. Note that the sensing dictionary given by the non-adaptive design method [12], which is completely determined by the dictionary $\mathbf{S}$ and independent of the received signal, corresponds to a special case of (19) with $\mathbf{U}^{(k)} = \mathbf{I}_L$ (an identity matrix) at each step of the general OMP algorithm.

The proposed algorithm is summarized as follows.
(1) Initialization: $\mathbf{g}_0 = \mathbf{r}$, $\mathbf{W} = \mathbf{S}$, $k = 0$;
(2) for $i = 0, 1, \cdots, L-1$, repeat the following process for $J$ times:

$$\hat{\mathbf{h}}^{(k)} = \left|\mathbf{W}^T \mathbf{g}_k\right|, \ \mathbf{U}^{(k)} = \text{diag}(\left|\hat{\mathbf{h}}^{(k)}\right|^\rho),$$

$$\mathbf{D}_i = \frac{1}{\mathbf{s}_i^T \left(\mathbf{S}\mathbf{U}^{(k)}\mathbf{S}^T + \beta \mathbf{I}_M\right)^{-1} \mathbf{s}_i} \left(\mathbf{S}\mathbf{U}^{(k)}\mathbf{S}^T + \beta \mathbf{I}_M\right)^{-1},$$

$$\mathbf{w}_i = \mathbf{D}_i \mathbf{s}_i, \ \mathbf{W} = [\mathbf{w}_0 \ \mathbf{w}_1 \ \cdots \ \mathbf{w}_{L-1}];$$

(3) $\hat{\mathbf{h}}^{(k)} = \left[\hat{h}_0^{(k)} \ \hat{h}_1^{(k)} \ \cdots \ \hat{h}_{L-1}^{(k)}\right]^T = \left|\mathbf{W}^T \mathbf{g}_k\right|$,

$$m_k = \arg \max_{0 \leq i \leq L-1} \hat{h}_i^{(k)},$$

$$\hat{\mathbf{A}}^{(k)} = \left[\mathbf{s}_{m_0} \ \mathbf{s}_{m_1} \ \cdots \ \mathbf{s}_{m_k}\right],$$

$$\mathbf{P}_k = \mathbf{I}_M - \hat{\mathbf{A}}^{(k)} \left(\hat{\mathbf{A}}^{(k)T} \hat{\mathbf{A}}^{(k)}\right)^{-1} \hat{\mathbf{A}}^{(k)T}, \ \mathbf{g}_{k+1} = \mathbf{P}_k \mathbf{r}.$$

(4) $k = k+1$, go to (2) and repeat until $k = K$.

Finally, the position of the nonzero entries of SMPC is detected by $[m_0 \ m_1 \ \cdots \ m_{K-1}]$, and the corresponding nonzero values are estimated as $\left(\hat{\mathbf{A}}^{(k)T}\hat{\mathbf{A}}^{(k)}\right)^{-1} \hat{\mathbf{A}}^{(k)T} \mathbf{r}$. To reduce the computation cost, the sensing dictionary can be calculated only for $k = 0$ and used at the subsequent steps.

V. SIMULATION RESULTS

To gain some insights into the effect of IAI mitigation on sparse channel estimation, we carried out 10000 independent Monte-Carlo trials. The nonzero entries of SMPC are drawn randomly from a uniform distribution on $[-1, -0.2] \cup [0.2, 1]$ and the number of nonzero entries is $K=5$. The position of nonzero entry of $\mathbf{h}$ is generated randomly. The channel length is set as $L=50$ or 100, the length of training sequence is $N=30$, and the signal to noise ratio (SNR) is 10 dB. The other involved parameters used in the algorithms are set to $\rho = 3$, $\alpha = 0.1$, $\beta = 0.1$, and $J = 10$, which may be further optimized to obtain better performance. Simulation results are obtained over we compare performance of the general OMP algorithm based on adaptive IAI mitigation (adaptive IAI) with that of the least squares method (LS), the ordinary OMP algorithm (OMP) and the general OMP algorithm based on non-adaptive IAI mitigation (non-adaptive IAI). As a benchmark, we also plot the results of the unrealistic IAI mitigation (12) using a prior information $\mathbf{B}$. First, we compare the ability of these algorithms to detect the nonzero entries of SMPC. The cumulative density functions (CDF) of the number of incorrectly detected nonzero components for the channel length values of 50 and 100 are shown in Figures 1 and 2, respectively. From these CDF functions, we see that the proposed general OMP algorithm based on adaptive IAI mitigation gives more accurate detection of nonzero entries of SMPC than other algorithms, especially in the latter sparser channel case. Second, we evaluate the ability of these algorithms to estimate correctly the nonzero entries of SMPC. The CDF of the estimation error of nonzero entries for the channel length values of 50 and 100 are illustrated in Figures 3 and 4, respectively. Here, the sensing dictionary is calculated only for $k = 0$ and used at the subsequent steps. Because the number of nonzero entries of SMPC is small, it has been observed that the performance loss is negligible. It can be seen that the proposed general OMP algorithm based on adaptive IAI mitigation is significantly better than the other three methods, especially in the case of sparser channel ($L$=100).

VI. CONCLUSION

In this paper, we present a novel inter-atom interference mitigation method for the general OMP algorithm to improve the performance of sparse multi-path channels estimation especially in the case of sparser and longer SMPC. The numeral experiments indicate that the proposed algorithm outperforms both the ordinary OMP algorithm and the general OMP algorithm based on non-adaptive IAI mitigation.

ACKNOWLEDGMENT

The authors thank anonymous reviewers for their valuable suggestions.

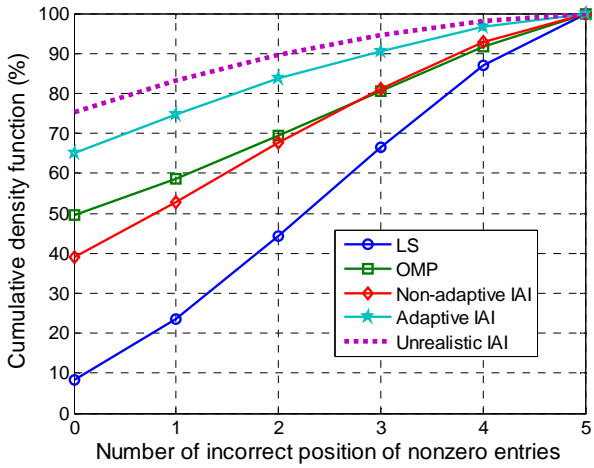

Fig.1 Cumulative density function of the number of incorrect position of nonzero entries. (N=30, L=50)

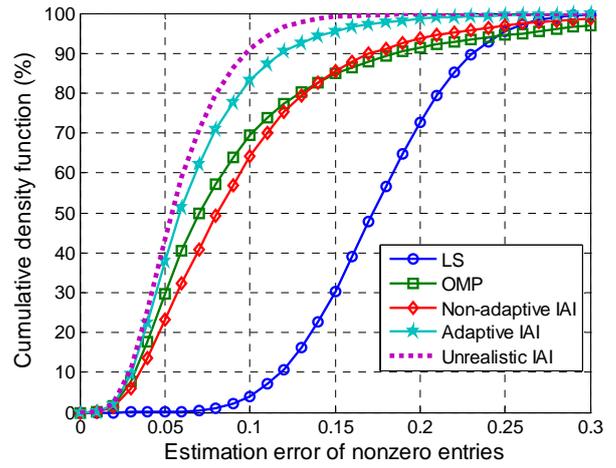

Fig.3 Cumulative density function of the estimation error of the nonzero entries (N=30, L=50).

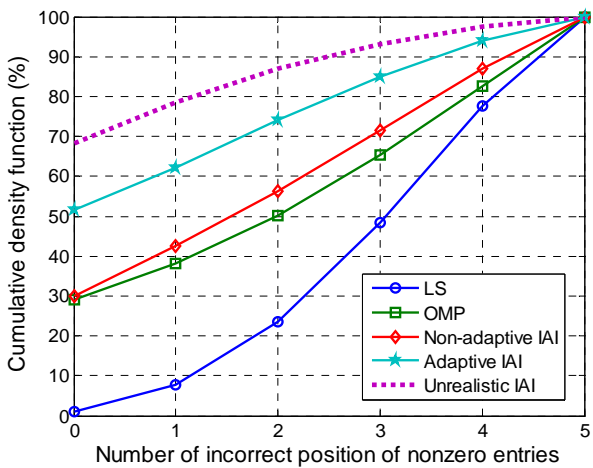

Fig.2 Cumulative density function of the number of incorrect position of nonzero entries (*N*=30, *L*=100).

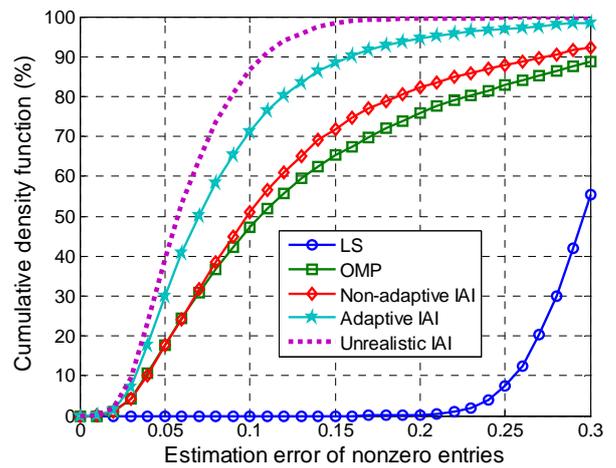

Fig. 4 Cumulative density function of the estimation error of the nonzero entries (N=30, L=100).